\documentclass{article}
\usepackage{spconf,amsmath,graphicx}
\usepackage{acronym,url}
\usepackage[hidelinks]{hyperref}
\usepackage{makecell}
\usepackage{enumitem}
\usepackage{xcolor}
\usepackage{booktabs}
\usepackage{flushend}
\usepackage{amssymb}

\usepackage{tikz}
\usepackage{pgfplots}
\usepackage{graphics}
\usetikzlibrary{positioning, calc}
\usepackage{placeins}

\acrodef{SASSEC}[SASSEC]{Stereo Audio Source Separation Evaluation Challenge}
\acrodef{BAQ}[BAQ]{Basic Audio Quality}
\acrodef{SiSEC}[SiSEC]{Signal Separation Evaluation Campaign}
\acrodef{SEBASS}[SEBASS]{Subjective Evaluation of Blind Audio Source Separation}
\acrodef{DNN}[DNN]{Deep Neural Network}
\acrodef{PSM}[PSM]{Phase-Sensitive Mask}
\acrodef{STFT}[STFT]{Short-Time Fourier Transform}
\acrodef{DS}[DS]{Dialogue Separation}
\acrodef{DE}[DE]{Dialogue Enhancement}
\acrodef{SD}[SD]{Standard Deviation}

\newcommand\numExpertsIIS{15}  % number of expert listeners from IIS before post-screening
\newcommand\numExpertsNFX{13}  % number of expert listeners from Netflix before post-screening
\newcommand\numExperts{26}  % number of expert listeners (after post-screening)
\newcommand\numScores{6240}  % number of subjective scores 
\newcommand\avgExp{12.7}  % average year of professional experience
\newcommand\avgAge{37.5}  % average age
\newcommand{\scale}{0.6}

\title{ODAQ: Open Dataset of Audio Quality}

\name{
\begin{tabular}{c}
Matteo Torcoli$^{1\star}$, Chih-Wei Wu$^{2\star}$, Sascha Dick$^{1\star}$, Phillip A. Williams$^{2\star}$,\\ 
Mhd Modar Halimeh$^1$, William Wolcott$^2$, Emanuël A. P. Habets$^1$
\thanks{$^\star$ Co-first authors.}
\end{tabular}
}
\address{$^1$Fraunhofer Institute for Integrated Circuits IIS, Erlangen, Germany \\ $^2$Netflix, Inc., Los Gatos, USA}

\usepackage[pscoord]{eso-pic}
\newcommand{\placetextbox}[3]{% \placetextbox{<horizontal pos>}{<vertical pos>}{<stuff>}
\setbox0=\hbox{#3}% Put <stuff> in a box
\AddToShipoutPictureFG*{% Add <stuff> to current page foreground
\put(\LenToUnit{#1\paperwidth},\LenToUnit{#2\paperheight}){\vtop{{\null}\makebox[0pt][c]{#3}}}%
}%
}%

\begin{document}
\ninept
\maketitle
\begin{abstract}
Research into the prediction and analysis of perceived audio quality is hampered by the scarcity of openly available datasets of audio signals accompanied by corresponding subjective quality scores. 
To address this problem, we present the Open Dataset of Audio Quality (ODAQ), a new dataset containing the results of a MUSHRA listening test conducted with expert listeners from 2 international laboratories. ODAQ contains 240 audio samples and corresponding quality scores. Each audio sample is rated by \numExperts\ listeners.
The audio samples are stereo audio signals sampled at 44.1 or 48\,kHz and are processed by a total of 6 method classes, each operating at different quality levels.
The processing method classes are designed to generate quality degradations possibly encountered during audio coding and source separation, and the quality levels for each method class span the entire quality range.
The diversity of the processing methods, the large span of quality levels, the high sampling frequency, and the pool of international listeners make ODAQ particularly suited for further research into subjective and objective audio quality.
The dataset is released with permissive licenses, and the software used to conduct the listening test is also made publicly available.
\end{abstract}
\begin{keywords}
Audio Quality, Subjective Scores, Openly Available, Dataset, Psychoacoustics
\end{keywords}

\placetextbox{0.5}{0.08}{\fbox{\parbox{\dimexpr\textwidth-2\fboxsep-2\fboxrule\relax}{\footnotesize \centering Accepted paper. \copyright  2023 IEEE. Personal use of this material is permitted. Permission from IEEE must be obtained for all other uses, in any current or future media, including reprinting/republishing this material for advertising or promotional purposes, creating new collective works, for resale or redistribution to servers or lists, or reuse of any copyrighted component of this work in other works.}}}

\section{Introduction}
\label{sec:intro}
To investigate questions related to perceived audio quality (subjective quality) and how to predict it (objective quality), researchers need datasets of audio signals and corresponding audio quality scores.
Gathering this type of data is a lengthy and costly process. Moreover, audio signals often cannot be shared due to copyright restrictions. As a consequence, many datasets of audio quality are kept private or shared only to a limited extent. This is observed frequently for datasets targeting audio-coding applications. For example, the USAC verification test~\cite{n12232} has been widely used (e.g., \cite{Torcoli2021, biswas2023}) % delgado2020
to analyze the correlation between objective measures and subjective scores.
However, the dataset is not publicly available, posing reproducibility issues and disadvantages for the researchers who cannot access it.
The few existing publicly available datasets are limited in size,  spanned quality range, scope (e.g., focusing exclusively on a single application domain), or quality of the considered audio signals (e.g., sampling frequency) or of the subjective scores (e.g., listening test carried out under uncontrolled conditions).

This work presents the Open Dataset of Audio Quality (ODAQ), a new dataset distributing both audio signals and subjective scores under permissive licenses.
The \numScores\ subjective scores are \ac{BAQ} scores collected via an extensive MUSHRA test~\cite{MUSHRA} involving \numExperts\ expert listeners.
The reference audio signals are collected with audio for broadcast and streaming in mind. These are studio-quality stereo signals that are then processed by two sets of audio processing methods representing 1) audio coding and 2)~source separation for \ac{DE}. Using processing methods from different application domains provides new insight into how audio quality is rated when different audio signal processing application domains are jointly considered in one listening test. This renders ODAQ a diverse validation set for objective measures. Diversity is important to benchmark objective measures as they often achieve reasonable performance for specific domains, but do not generalize well out of them \cite{Torcoli2021, wu2021}. 
The ODAQ package is made available online.\footnote{\url{https://github.com/Fraunhofer-IIS/ODAQ/}}\textsuperscript{,}\footnote{\url{https://doi.org/10.5281/zenodo.10405774}}

\section{Existing Datasets}
This section reviews existing datasets of audio signals and subjective scores described in literature. 
Focusing on audio coding, \ac{BAQ} ratings~\cite{MUSHRA} are obtained in \cite{dick2017} to evaluate the quality degradations caused by isolated coding artifacts. 
The subjective scores from \cite{dick2017} are used to investigate and develop objective measures~\cite{Torcoli2021, delgado2022data}. However, the material could not be shared openly due to copyright restrictions. 
Motivated also by this, we decided to employ a similar approach as in \cite{dick2017}, and to expand it to material that can be shared, along with re-tuned artifacts, a larger set of processing methods (i.e., including \ac{DE}), and a larger and more diverse cohort of listeners.

Meanwhile, \cite{n12232, n10032, n13633, itu_bs1387} utilize listening tests to evaluate audio coding methods or objective measures for this application. These datasets are, however, not fully publicly available. 
In contrast, other datasets on audio coding quality are openly available \cite{Martinez14, multiformat_lt}, but they are either small in size \cite{Martinez14} or the listening tests are conducted under uncontrolled conditions~\cite{multiformat_lt}. 
Finally, the EBU SQAM CD~\cite{EBUSQAM} is often used as audio material for testing coding systems, and it was recently used for the educational package in \cite{herre2023}. However, the SQAM CD is made available with license restrictions.

In the field of source separation, PEASS \cite{Emiya2011} is one of the first publicly available datasets of audio signals and relative scores, where four source separation algorithms are considered. Moreover, \mbox{SEBASS}~\cite{Kastner2022} comprises \ac{BAQ} scores for audio signals processed by a larger variety of source separation algorithms, and it includes audio signals from different openly-available sources. The audio signals in PEASS and SEBASS can be generally described as speech signals sampled at $16$~kHz and music signals sampled at $16$ or $44.1$~kHz.

Distortions typically introduced in telecommunication applications are the focus of NISQA \cite{Mittag21} which, in addition to the predictive model weights, provides datasets with $48$~kHz audio signals
and their subjective ratings. The provided ratings span a number of speech quality dimensions, such as noisiness and coloration, as well as overall speech quality. This is expanded with more datasets focusing on online conferencing in \cite{yi2022}.

The authors in \cite{Cutler22} describe large datasets of listening test results obtained with audio signals 
processed by different noise suppression algorithms. The results are then used to develop non-intrusive, \ac{DNN}-based quality metrics. 
However, the results of the subjective listening tests are not publicly available.

Finally, recent publications have investigated the subjective quality of synthesized speech signals (an application that is beyond the scope of this paper) and provided publicly available datasets for this purpose, e.g., \cite{Maniati22}. 

\section{Test Material}
\label{sec:odaq}

\subsection{Audio Material}

The audio material originates from different sources and includes both samples already openly available on the web and newly released productions by Netflix and Fraunhofer IIS. The material can be downloaded, modified, and redistributed under the licenses specified in the ODAQ package. 
As original material, only studio-quality audio sampled at 44.1 or 48~kHz is considered. 
The material is carefully selected with the goal of representing broadcast and streaming audio, and to then generate a large range of degradations and quality levels after processing with the selected methods.  Table~\ref{tbl:artifacts} gives an overview of the considered processing methods and quality levels, which are explained in Sec.\,\ref{sec:aca} and Sec.\,\ref{sec:ds}.

The original material consists of 11 movie-like soundtracks with dialogues mixed with music and effects (separate stems and transcripts are also provided) and 14 music excerpts (8 of which are solo recordings).
In particular, the music items are processed by 5 methods simulating isolated artifacts potentially originating from audio coding at different quality levels, as detailed in Sec.\,\ref{sec:aca}.
Two of the music items are presented processed by all 5 audio-coding-related methods to allow the comparison (left for future work) of different degradations on the same audio content.
The movie-like soundtracks are processed by 5 different systems performing source separation for \ac{DE}, as discussed in Sec.\,\ref{sec:ds}.
In the case of DE, a mix of degradations can appear simultaneously. 

Without considering the training session (the audio material for this phase is also available in the ODAQ package), ODAQ consists of 30 trials; each trial consists of one source (i.e., reference), two low-pass anchors as recommended by \cite{MUSHRA}, and 5 processed samples. For each trial, the 5 processed samples correspond to one row of Table~\ref{tbl:artifacts}. This results in a total number of 240 audio samples with an average duration of 12.5 seconds.

As different trials can present different types of degradations, an additional challenge is posed for the participants in the listening test, who are asked to score degradations potentially sounding very differently, one trial after the other.
Our intention is to minimize the centering bias \cite{zielinski:2008}, 
encourage the listeners to anchor their judgment on the absolute quality score,
and to build a dataset that can represent perceived audio quality beyond a specific degradation or application domain \cite{Torcoli2021}.

\begin{table}
\begin{footnotesize}
\centering
\setlength{\tabcolsep}{4pt}
\begin{tabular}{c c c c c c c}
\toprule
\textbf{Processing} & \textbf{Parameter} & \multicolumn{5}{c}{\textbf{Quality Level}} \\[2pt] \cmidrule{3-7}
\textbf{} & \textbf{} & \textbf{Q1} & \textbf{Q2} & \textbf{Q3} & \textbf{Q4} & \textbf{Q5}\\ \midrule
LP & freq. [kHz]  & $5.0$ & $9.0$ & $10.5$ & $12.0$ & $15$ \\ 
TM & freq. [kHz] & $3.0$ & $5.0$ & $7.0$ & $9$ & $10.5$ \\ 
UN & freq. [kHz] & $3.0$ & $5.0$ & $7.0$ & $9.0$ & $10.5$ \\ 
SH & hole prob. [\%] & $70$ & $50$ & $30$ & $20$ & $10$ \\ 
PE & \makecell{NMR[db]\\length}& \makecell{10\\4096}  & \makecell{10\\2048} &  \makecell{10\\1024} & \makecell{16\\2048} & \makecell{16\\1024} \\ 
DE & DS system & \cite{openunmix} & \cite{TFCTDF} & \cite{Petermann22} & \cite{DeepFilterNet2} & q-PSM \\ \bottomrule
\end{tabular}
\caption{\label{tbl:artifacts}Processing methods and quality levels used to generate the conditions under test.}
\label{TableArtifactParameters}
\end{footnotesize}
\end{table}

\subsection{Audio Coding Artifacts}
\label{sec:aca}

The work in \cite{dick2017} presents a method to generate isolated audio coding artifacts by forcing audio coders into controlled, sub-optimal operation modes. 
The following 5 types of artifacts are generated: \\
\textbf{1) Low-Pass (LP)}, referred to as Bandwidth Limitation (BL) in \cite{dick2017}. This is generated the same way the low-pass MUSHRA anchors are generated. This means that, with respect to Table~\ref{tbl:artifacts}, two additional levels (3.5 and 7 kHz) are present for this degradation.\\
\textbf{2) Tonality Mismatch (TM)} refers to distortions resulting from noise-like components being substituted with tonal components, e.g., in bandwidth extension when using a scaled copy of the lower part of the spectrum for high frequencies. \\
\textbf{3) Unmasked Noise (UN)} refers to distortions resulting from tonal components being substituted by noise with the same spectral envelope, e.g., in parametric coding or bandwidth extension. \\
\textbf{4) Spectral Holes (SH)}: spectral parts are quantized to zero, e.g., due to coarse quantization in transform coders. Here, this is simulated by quantizing bands to zero with a controlled hole probability.\\
\textbf{5) Pre-Echoes (PE)} refer to temporal smearing of quantization noise around transients, e.g., in filterbank-based coding approaches.

This selection comprises well-known artifacts that are typical for waveform-preserving audio coders (LP, SH, PE). Furthermore, it includes degradations that are specific to more recent generations of parametric audio coders (TM, UN), which are generally underrepresented in existing datasets, especially those predating the widespread use of parametric coding techniques.

To refine the uniformity of the distribution of the quality levels, the parametrization of the artifact generation (detailed in Table \ref{tbl:artifacts}) is adjusted with respect to \cite{dick2017}. The adjustments are based on the conclusions from  \cite{dick2017} and on preliminary listening sessions carried out by the authors.
For LP, TM, and UN, the cross-over frequencies are adapted to be more evenly spaced in the lower frequency range.
The generation of PE is modified to enable finer control of the quality levels by separately controlling the Noise-to-Mask Ratio (NMR) \cite{brandenburg1987evaluation} as well the transform block length. This allows to vary quality via the \emph{amount} of temporally smeared quantization noise around transients (as used for the quality levels in \cite{dick2017}), and additionally variation of quality by changing the \emph{temporal extent} of the smearing while keeping the same NMR.
Furthermore, the generation of SH and PE artifacts is extended by applying Mid-Side (MS)-Stereo processing to mitigate the introduction of additional spatial artifacts.
 The Mid and Side signals are both individually processed by the respective artifact generation methods and then transformed back into left and right signals. 

All items were normalized to $-23$\,LUFS integrated loudness~\cite{ebu_r128} before processing. The reference signal for the listening test is the original material before processing.

\subsection{Dialogue Separation (DS) and Remixing}
\label{sec:ds}

Given an audio mixture $y_i(n) = s_i(n) + v_i(n)$, typically stereo, i.e., $i \in \{1,2\}$, the \ac{DS} task \cite{torcoli2023}
is to estimate the foreground speech signal $s_i(n)$, also referred to as dialogue, and separate it from $v_i(n)$, which denotes the background components such as music, effects, and/or noise. In this work, we consider dialogue in the phantom center of the stereo scene. In particular, to avoid spatial artifacts the centered mixture signal $y(n) = \frac{1}{2} \sum_{i=1}^{2} y_i(n)$ is used as input to the \ac{DS} methods, and the output dialogue estimate is made double mono.
The mixtures are processed by an oracle-knowledge approach and 4 state-of-the-art masking-based separation methods, selected by the authors via preliminary informal listening tests with the goal of uniformly spanning the perceived quality scale.

The oracle-knowledge approach is referred to as q-PSM, and employs a quantized \ac{PSM} in the \acl{STFT} domain, using signal frames of $2048$ samples with a $50\%$ overlap. The \ac{PSM} is calculated per time-frequency bin as described in \cite{Erdogan15} using the ground-truth dialogue and background signals.
To induce clearer perceived differences to the ground-truth reference dialogue, the \ac{PSM} mask is quantized uniformly with a quantization stepsize $\gamma=0.5$ such that for each time-frequency bin $(\tau,f)$, the quantized mask is defined as
\begin{equation}
    \mathcal{M}_{\textrm{q-PSM}}(\tau, f)  = \left\lfloor \frac{\mathcal{M}_{\textrm{PSM}}(\tau, f)}{\gamma} \right\rceil \, \cdot \, \gamma; \; \mathcal{M}_{\textrm{PSM}}(\tau, f) \in [-1, 1]
\end{equation}
where $\lfloor \cdot \rceil$ denotes the rounding to the nearest integer operator. 

The 4 state-of-the-art separation methods are based on \acp{DNN} and are evaluated using the implementations and pre-trained networks made publicly available by the original authors. These methods are 
Open Unmix \cite{openunmix},
TFC-TDF Unet \cite{TFCTDF},
Cocktail Fork Baseline Model \cite{Petermann22}, and
\mbox{DeepFilterNet 2}~\cite{DeepFilterNet2}.
Most of the selected methods were not originally designed for dialogue separation. The selection was motivated by their suitability of representing source separation artifacts at different quality levels.
The exact versions and the links to the repositories are reported in the ODAQ package.

Finally, the estimated dialogue $\hat{s}_i(n)$ are re-mixed with the corresponding background estimates $\hat{v}_i = y_i(n) - \hat{s}_i(n)$, rendering the \ac{DE} mixtures
\begin{equation}
\label{eq:remix}
    y_{\textrm{DE},i}(n) = \hat{s}_i(n) + \alpha \, \hat{v}_i(n), 
\end{equation}
where $\alpha$ corresponds to $-20$~dB. For the listening test, the reference signal is obtained by remixing the ground-truth signals, i.e., using $s_i(n)$ and $v_i(n)$ instead of their estimates in Eq.\,\ref{eq:remix}. The \ac{DE} stereo mixtures are normalized to $-23$~LUFS integrated loudness \cite{ebu_r128}. 

\section{Listening Test}
\subsection{Test Method and Setup}
\label{sec:lt}
The MUSHRA test was carried out in 2 different laboratories based in Germany and the United States (US): 
\numExpertsIIS\ listeners participated in Germany, while \numExpertsNFX\ listeners participated in the US. A post-screening procedure was conducted, which excluded the results for 1 listener from each laboratory for having rated the hidden reference below 90 MUSHRA points for more than 15\% of all trials \cite{MUSHRA}.

The remaining \numExperts\ listeners have a mean age of \avgAge\ years (\ac{SD}: $8.5 $), have no hearing impairments based on self-reports, and have backgrounds in areas such as audio mixing/production, music performance, or audio research/development; the average years of professional audio experience is \avgExp\ (\ac{SD}: $10$). 
Both laboratories have an international subject pool, with a total of $6$ first languages represented in the pool of participants.
Still, the majority of the participants have either German or English as first language, depending on the laboratory.
All tests were conducted in acoustically damped listening rooms that are suitable for audio mixing or screening. The audio playback device was Beyerdynamic DT770 Pro 250 Ohm closed-back headphones connected to a professional soundcard and a computer.

A training session, which was designed to familiarize the listeners with the methodology, user interface, and the test conditions, was provided to each listener prior to the full test. This training session includes 3 trials with audio samples that are not used in the full test. The listeners were instructed to adjust the playback to a comfortable level during training and avoid further adjustments in the full test. A written instruction in English or German, which contained the detailed information of the procedure, was made available to the listeners, and doubts were clarified verbally.

The full test was divided into 3 sessions with 10 trials each; listeners were strongly advised to take long breaks between sessions to reduce fatigue. Most of the listeners took the 3 sessions on different days. In total, \numScores\ subjective scores were collected. All the post-screened listening test results, including the raw scores and corresponding metadata, are included in the ODAQ package.  

\begin{figure}[tb]
\centering
\centerline{\includegraphics[width=1\linewidth]{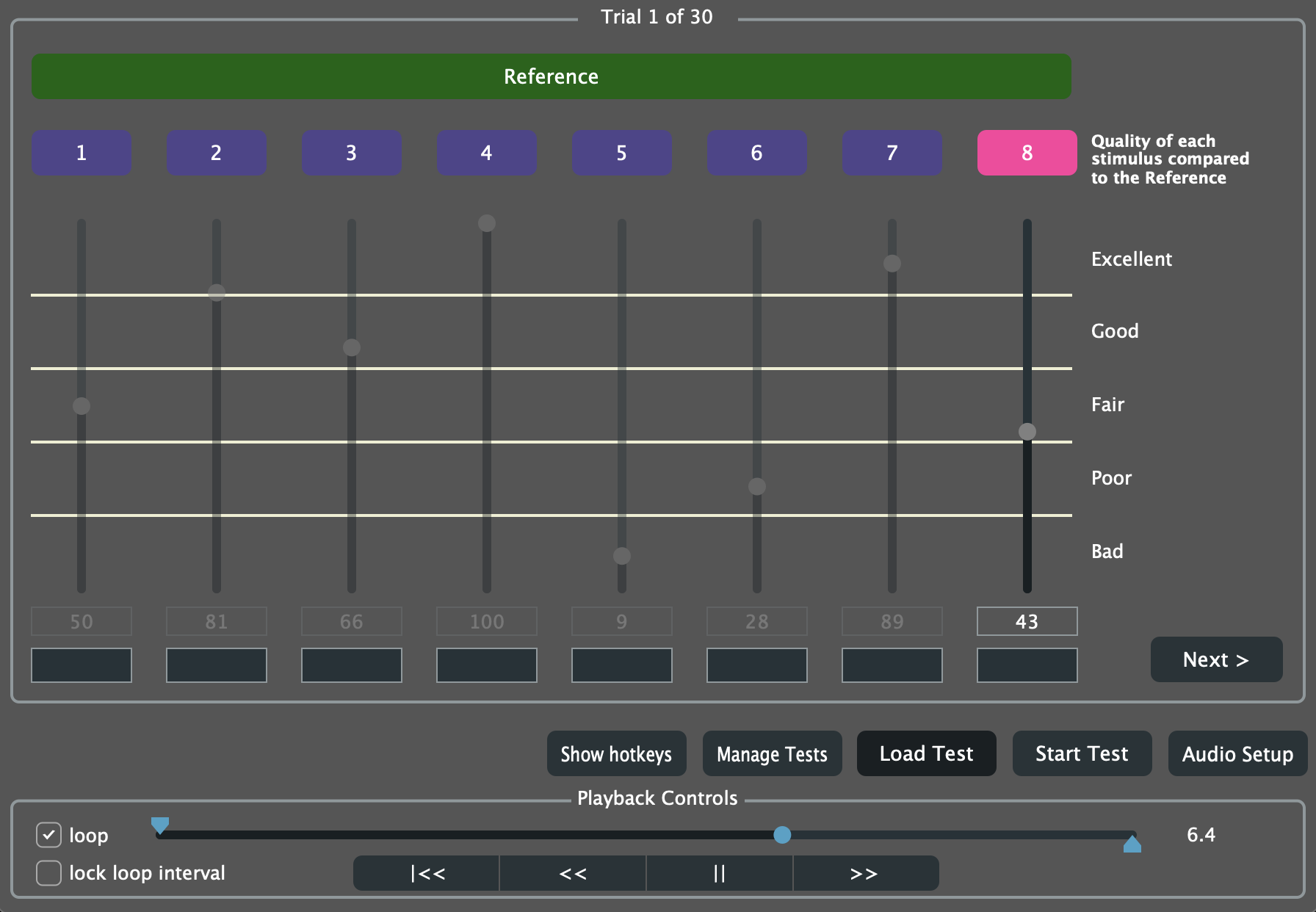}}
%  \vspace{1.5cm}
\caption{{\label{fig:GUI}}Screenshot of the listening test user interface.}
\end{figure}

\subsection{Software}
All the listening tests were conducted using the same custom-built software tool.
The listening test application, built on top of the JUCE\footnote{\url{https://juce.com/}
%, last accessed: 2023/08/30
} framework, supports multiple standard subjective listening tests procedures
and provides the foundation for future extensions. The application implements basic user interface components shown in English and essential audio playback features such as instant sample switching and looping (a screenshot is provided in Figure~\ref{fig:GUI}), and it does not require an internet connection. 
The tool also implements the splitting of the trials in multiple sessions, so to easily preserve the randomization of the trials across the full test, instruct the listener to take a break at the right point in the test, and allow to easily continue the test from where it was interrupted.
The source code of the application, including a compiled version (v3.2.2) which is compatible with macOS version 10.13 and later, is available online.\footnote{\url{https://github.com/Netflix-Skunkworks/listening-test-app}}

\begin{figure*}[htb]
\centering
\centerline{\includegraphics[width=\linewidth,trim=0cm 0.4cm 0cm 0.45cm, clip=true,]{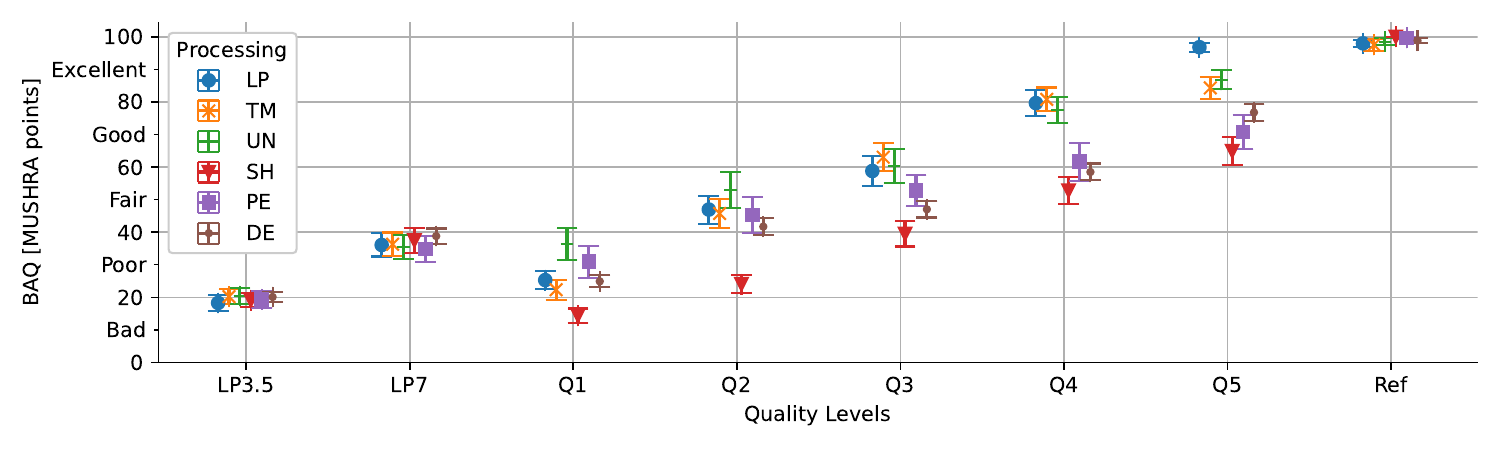}}
\vspace{-0.3cm}
\caption{{\label{fig:results}}Overall results (\numExperts\ participants) showing mean \acf{BAQ} scores and 95\% confidence intervals (CI) per processing method and quality level (see Table \ref{tbl:artifacts}) as well as low-pass anchors and hidden reference. The horizontal spacing between the processing methods at each quality level is intended solely for better visibility. Note that the hidden reference (Ref) and the low-pass anchors (LP3.5 and LP7) are depicted with different colors and symbols depending on which processing method was presented along with them in the trial.}
\end{figure*}

\begin{figure}[htb]
\centering
\centerline{\includegraphics[width=0.8\linewidth]{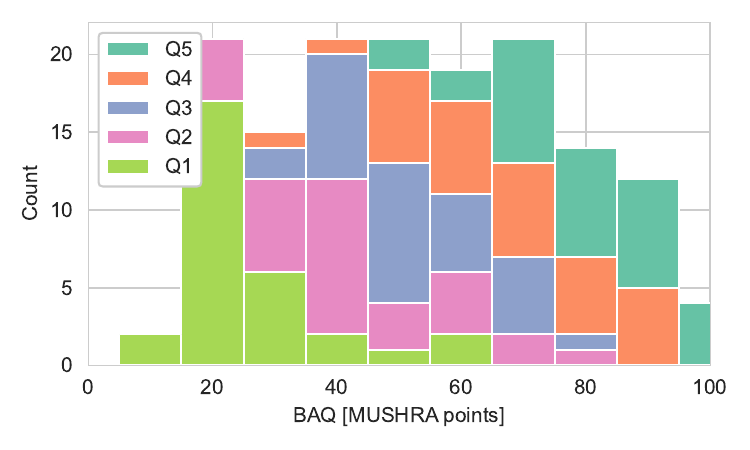}}
\vspace{-0.3cm}
\caption{{\label{fig:distro}}Histogram of mean \ac{BAQ} scores, excluding Ref/Anchors. 
}
\end{figure}

\begin{figure}[htb]
    \centering
    \input{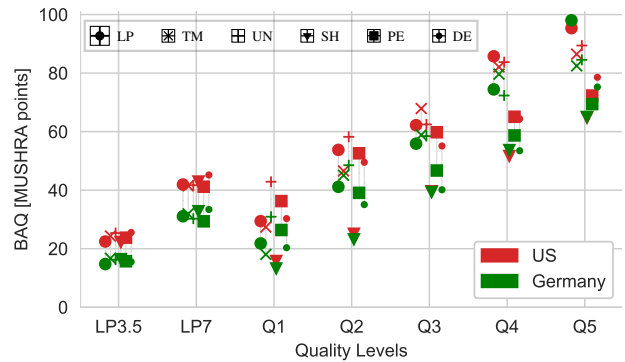}
\vspace{-0.3cm}
\caption{{\label{fig:labs}}Mean \ac{BAQ} scores, separated per laboratory, same order and markers used as in Figure~\ref{fig:results}. A general offset between the two laboratories can be observed, while the results for some conditions match beyond expectation, e.g., SH (depicted by $\blacktriangledown$).}
\end{figure}

\section{Listening test results}
\label{sec:results}

\vspace{-0.2em}
Figure~\ref{fig:results} shows the overall results from the listening test as mean \ac{BAQ} scores and 95\% confidence intervals (CI). 
As also shown in Figure~\ref{fig:distro}, the mean scores span the most relevant portion of the MUSHRA scale, from \textit{Poor} to \textit{Excellent}. 
As also reported in \cite{zielinski:2008}, listeners tend to avoid using the extremes of the scale (contraction bias). While some zeros are observed in the raw data, the lowest mean score is $10.6$, registered for one item processed with SH in Q1. 
On the other extreme, most of the scores above $95$ refer to the hidden reference, with the exception of LP in Q5, which approaches transparency. 
The range of quality between $15$ and $95$ MUSHRA points is fully spanned with at least $12$ mean scores in each 10-point interval (Figure~\ref{fig:distro}).
In addition, each portion of the 15-85 range is spanned by multiple processing methods. For example, scores close to $60$ MUSHRA points are obtained for LP, TM, and UN in Q3 as well as for PE and DE in Q4 (Figure~\ref{fig:results}).
The large quality range and the overlap of methods at similar perceived quality levels make these scores particularly useful as benchmark for objective measures, and to assess their generalization capabilities.

Figure~\ref{fig:labs} shows the difference in mean scores between the two laboratories. 
A general offset can be observed, with the quality scores from the US being generally higher than the results from Germany, e.g., approximately 6 to 12 MUSHRA points for the low-pass anchors,
and differences becoming smaller at higher quality levels (especially for Q4 and Q5). SH is an exception to this trend, with the mean scores matching almost perfectly across laboratories for all quality levels.
The general offset might be explained by a different linguistic attitude towards the quality labels \cite{zielinski:2008}. 
This highlights the importance of pooling together the results from different regions of the world to better sample the quality perception of the world population, as also reported by \cite{Maniati22}.

\vspace{-0.2em}
\section{Conclusions}
\label{sec:conclusions}

\vspace{-0.2em}
This work presented the Open Dataset of Audio Quality (ODAQ), a dataset of audio signals and subjective scores distributed under permissive licenses. The audio signals are distributed as uncompressed stereo audio at $44.1$ or $48$\,kHz.
The subjective scores are collected via a MUSHRA test with \numExperts\ expert listeners (after post-screening) from two international laboratories.

The diversity of the processing methods, of the considered application domains, of the quality levels, and the international nature of the participant pool make this dataset particularly suited for further research in the fields of subjective and objective audio quality.
Further analysis of the subjective results as well as benchmarking state-of-the-art objective measures are natural follow-up works, left for the future and for the community.

\section{Acknowledgment}
\label{sec:ack}
The authors warmly thank the participants in the listening test.
\clearpage

\vfill\pagebreak

\bibliographystyle{IEEEbib}
\bibliography{refs}

\end{document}